\documentclass[11pt]{llncs}
\usepackage{llncsdoc}
\usepackage{amsfonts,amsmath,amssymb,latexsym}
\ifx\pdftexversion\undefined
  \usepackage[colorlinks,linkcolor=black,filecolor=black,citecolor=black,urlcolor=black,pdfstartview=FitH]{hyperref}
\else
   \usepackage[colorlinks,linkcolor=blue,filecolor=blue,citecolor=blue,urlcolor=blue,pdfstartview=FitH]{hyperref}

\usepackage{graphicx}
\usepackage{xspace}
\usepackage{subfigure}
\usepackage{lscape}
\usepackage{color}
\newcommand{\minH}{\mathtt{minH}}
\newcommand{\minHG}{\mathtt{minHG}}
\newcommand{\ZH}{\mathtt{bestH}}

\newcommand{\cleanr}{r_c}
\newcommand{\FS}{\mathtt{FS}}
\newcommand{\go}{\mbox{\textsc{go}}\xspace}
\newcommand{\concon}{\mbox{\textsc{ConCon}}\xspace}
\newcommand{\ccfs}{\mbox{\textsc{CCfs}}\/}

\newcommand{\ie}{\emph{i.e.,\ }}
\newcommand{\Ie}{\emph{I.e.,\ }}

\newcommand{\namedref}[2]{\hyperref[#2]{#1~\ref*{#2}}}
\newcommand{\sectionref}[1]{\namedref{Section}{#1}}

\newcommand{\theoremref}[1]{\namedref{Theorem}{#1}}

\newcommand{\figureref}[1]{\namedref{Figure}{#1}}

\newcommand{\lemmaref}[1]{\namedref{Lemma}{#1}}

\newcommand{\definitionref}[1]{\namedref{Definition}{#1}}

\newcommand{\hk}{{k''}}
\newcommand{\stab}{{\mathtt{stab}}}
\newcommand{\hide}[1]{}

\newcommand{\dd}[1]{{\sffamily\bfseries\color{red}\mbox{[[[DD:\ } #1 \mbox{]]] }}}

\newcommand{\ym}[1]{{\sffamily\bfseries\color{green}\mbox{[[[YM:\ } #1 \mbox{]]]  }}}
\newcommand{\ezra}[1]{\textbf{\color{blue}[[[EZRA: #1]]]}}
\newcommand{\tb}{\makebox[0.4cm]{}}
\newcommand{\due}{\makebox[0.8cm]{}}

\newcommand\A{\mathcal{A}}
\newcommand\BB{\mathcal{B}}

\newcommand\I{\mathcal{I}}
\newcommand\OO{\mathcal{O}}

\newcommand\p{\mathcal{P}}

\newcommand\F{\mathcal{F}}
\newcommand\s{\mathcal{S}}

\newcommand\x{\delta}
\newcommand\rh{\mathtt{disH}}
\newcommand\ah{\mathtt{absH}}
\newcommand\bb{\pi}

\newsavebox{\theorembox}
\newsavebox{\lemmabox}
\newsavebox{\conjecturebox}
\newsavebox{\claimbox}
\newsavebox{\factbox}
\newsavebox{\corollarybox}

\newsavebox{\propositionbox}
\newsavebox{\examplebox}
\savebox{\theorembox}{\bf Theorem}

\savebox{\lemmabox}{\bf Lemma}

\savebox{\conjecturebox}{\bf Conjecture}

\savebox{\claimbox}{\bf Claim} \savebox{\factbox}{\bf Fact}

\savebox{\corollarybox}{\bf Corollary}

\savebox{\propositionbox}{\bf Proposition}

\savebox{\examplebox}{\bf Example}
\newtheorem{notation}{{\sc Notation}\rm }
\newtheorem{observation}{{\sc Observation}\rm }

\newcommand{\ignore}[1]{}
\newcommand{\byzantine}[0]{\mbox{\emph{Byzantin}\hspace{-0.15em}\emph{e}}\xspace}
\newcommand{\safety}{\textit{``safety''}\xspace}
\newcommand{\simul}{\textit{``simultaneity''}\xspace}
\newcommand{\liveness}{\textit{``liveness''}\xspace}

\newcommand{\fireAlg}{\textsc{Fire-Squad}\xspace}

\newcommand\req{\mbox{\it Requests}}
\newcommand\fail{\mbox{\it Failed}\ \!\!}
\newcommand\failp{\mbox{{\it Failed}\ \!$'$}}
\newcommand\view{\mbox{\it Views}}
\newcommand\horz{\mbox{\it Horizon}}

\def\squarebox#1{\hbox to #1{\hfill\vbox to #1{\vfill}}}
\newcommand{\G}{\mathsf{G}}

\newcounter{linenumbers}
\newcommand{\linenumberspecific}[1]{{\tt #1}}
\newcommand{\linenumber}{\stepcounter{linenumbers}\linenumberspecific{\arabic{linenumbers}:}}
\newcommand{\lineref}[1]{Line~{\tt #1}}

\begin{document}

\title{An Optimal Self-Stabilizing Firing Squad}
\author{Danny Dolev\inst{1},\thanks{This research was supported in part by Israeli Science Foundation (ISF) Grant number 0397373.} Ezra N. Hoch\inst{1}, Yoram Moses\inst{2}}

\institute{
The Hebrew University of Jerusalem\\
Jerusalem, Israel\\
\and
Technion---Israel Institute of Technology\\
Haifa, Israel\\
}

\maketitle

\begin{abstract}
%\ezra{Author may improve the quality of the references by citing other works. Actually for 2/3 of the references there is at most an author in common.}
Consider a fully connected network where up to $t$ processes may
crash, and all processes start in an arbitrary memory
state. The self-stabilizing firing squad problem consists
of eventually guaranteeing simultaneous response to an external
input. This is modeled by requiring that the non-crashed processes ``fire''
simultaneously if some correct process received an external
%ezra1: of -> to
``$\go$''~input, and that they only fire as a response to some process
receiving such an input.
This paper presents \fireAlg, the first self-stabilizing firing squad algorithm. \\

The \fireAlg algorithm is optimal in two respects:
(a) Once the algorithm is in a safe state, it fires in response to
a $\go$ input as fast as any other algorithm does, and
(b) Starting from an arbitrary state, it converges to a safe state
as fast as any other algorithm does.
\end{abstract}

\section{Introduction}
The firing squad problem was first introduced in
\cite{BurnsLynch,DBLP:journals/siamcomp/CoanDDS89}. Informally, it is assumed that at any given round
a process may receive an external ``$\go$'' input, which is considered a
request for the
%yoram2
correct processes
%firing squad
to simultaneously ``fire.''
Roughly, a good solution is a protocol satisfying three
properties: (a) if some process fires in round $r$ then all
the non-crashed processes fire simultaneously in round~$r$;
(b) if a correct process receives a $\go$~input in round $r'$ then it
will fire at some later round $r> r'$; and
(c) a process fires in round $r$ only if
%ezra1: added footnote
%ezra1: \leq -> <
some process received a $\go$~input in some round $r' < r$.
(The formal definition disallows a solution in which a single input induces a constant firing.)

%ezra1: removed
%\dd{the properties allow a solution in which once anyone has an input you fire constantly}
%\ezra{I think that since this part is in the intro, and talks about BurnsLynch,DBLP:journals/siamcomp/CoanDDS89, it is ok that we don't give the precise def for SS-Byz-FireSqaud yet.}

Requiring the processes to fire simultaneously captures an important
aspect of distributed systems: There are cases in which it is
important that activities begin in the same round, {\it e.g.,} when
one distributed algorithm ends and another one begins, and the two may
interfere with each other if executed concurrently. Similarly, many
synchronous algorithms are designed assuming that all sites start
participating in the same round of communication. Finally,
simultaneity may be motivated by the fact that a distributed system
interacts with the outside world, and these interactions should often
be simultaneously consistent. A non-simultaneous announcement to
financial (stock) markets may enable unfair arbitrage trading, for example.

Coordinating simultaneous actions is not subsumed by the consensus
task. Indeed, even  when no transient failures are considered
possible (so there is a global clock and no self-stabilization is
required), solving the firing squad problem or simultaneously deciding
in a consensus task can be considerably harder than plain consensus
\cite{citeulike:4627304,DM}.
%yoram2 why does this follow?
This implies, in particular, that clock
%ezra1: added citations
synchronization \cite{citeulike:4627660,DigiClock-SSS06,DolWelSSBYZCS04,Lamport:1985:SCP,shamir-clcks} does not suffice for solving the
firing squad problem in a self-stabilizing manner; as it can be seen as providing round-numbers
to a self-stabilizing environment, which still leaves the firing squad problem as
a non-trivial problem.

\hide{\ym{Combine this with the clock-synch discussion;
emphasis is that simultaneity is a separate concern.}\ezra{changed the pgrph. Hope
it is better. If not, we can just remove it all together}
\ym{I think that we can cut if out. It is pretty much subsumed by the
previous and later ones.}
More specifically, when comparing simultaneous agreement to
the consensus problem, it turns out that there are different lower bounds
regarding the time it takes to solve each problem.
Simultaneous agreement depends solely on the failure pattern,
\ie on the order and times in which the
faulty processes stop operating (see \cite{DM,MT,MM}), while
there are non-simultaneous agreement solutions that
can terminate earlier depending on the input pattern,
\ie what is the input to each process in
every round (see
\cite{citeulike:4627304}, \cite{citeulike:4627277}). This implies that
simultaneity is a separate (and harder) problem than standard consensus.}

The firing squad problem is a primary example of a problem requiring
%ezra: nonfaulty->non-faulty (also in other places)
%yoram2
simultaneously coordinated
%simultaneous
actions by the non-faulty processes. Simultaneous
%yoram2
coordination has
%actions have
been shown to be closely related to the notion of common
knowledge \cite{HM,FHMV}, and this connection has been used to
characterize the
%yoram2 added
earliest
time required to reach simultaneous consensus, firing
squad, and related problems in a variety of failure models
\cite{DM,MT,NB,NT,MM,MR}. One of the consequences of this
literature is the
fact that the time at which
a simultaneous action that is based on initial values or external
inputs can be performed depends in a crucial way on the pattern in
which failures occur.

A general form of simultaneous agreement
called {\em continuous consensus} was defined in \cite{MM}.
In this problem, each of the processes maintains a list of events of
interest that have taken place in the run, and it is guaranteed that
the lists at all non-faulty processes are identical at all times.
They present an optimal
(non-stabilizing) implementation of such a service, which is a
protocol called \concon.
%XXX This service can
If we define as the events to be monitored by
\concon to be of the form $(\go,p,k)$, corresponding to
a $\go$ message arriving at process~$p$ at the end of round~$k$,
then a firing squad protocol can be obtained from \concon
simply by having the non-faulty processes fire exactly when a $(\go,p,k)$
event first appears in their identical copies of the ``common'' list.
We shall refer to this solution
to the firing squad problem
%yoram2 added
based on \concon
by $\ccfs$.

Traditionally, the firing squad problem assumes that processes do not
recover, \ie failed
processes stay failed forever. Moreover, even though it is easy to extend
the firing squad problem so that it can be repeatedly executed (\ie
allow for multiple firings over time, given that multiple $\go$~inputs are
received), it assumes that nothing in the system goes amiss---except
%yoram2 added
possibly
for the crash failures being accounted for. Adding
support for handling transient faults increases the robustness of  a
firing squad algorithm in this aspect. Indeed, a self stabilizing
solution will, in particular, be able to cope with
process recovery: Following process recoveries, the system will
%yoram2
eventually converge to a valid state and  continue operating correctly.
%converge to a valid state and will eventually continue operating correctly.

Transient faults alter a process's memory state in an arbitrary way.
A self-stabilizing algorithm \cite{DolevSSBook} is
assumed to start in an arbitrary state and be guaranteed to
eventually reach a state from which it operates according to its
intended specification.
Starting the operation at an arbitrary state enables the adversary to
``plant'' false information, such as the receipt of \go messages in
the past, which can cause the algorithm to  unjustifiably fire, either
immediately, or within a few rounds.
One of the challenges in designing an efficient self-stabilizing
firing squad algorithm is in bounding the damage that can be caused by
such false information in the initial state.

Perhaps the first candidate solution would be to
initiate an instance of $\ccfs$ in every round, with $t+1$ instances
executing concurrently at any given time, where $t$ is an upper bound
on the number of
%yoram2
possible crashed
%potentially crashing
processes. Firing would then take place
if it is
%yoram2
dictated by
%does in
any of the instances.
Since the component instances of such a solution are not themselves
stabilizing, all we can show is that such a solution is guaranteed to
stabilize after $t+1$ rounds, regardless of the failure pattern.
%yoram2
We shall present a solution that
%Our solution
does not consist of such a concurrent composition.
Moreover,
%yoram2
it performs
%we perform
subtle consistency checks
%yoram2
%comparing information
%at distinct processes, as well as a consistency check regarding
%reports of past failures,
to restrict the impact of
false information that appears in the initial state.
As a result, in some cases we obtain stabilization in as little as two
rounds.

The above discussion points out  the stabilization time as
an important aspect of a self-stabilizing firing squad algorithm.
Another central performance parameter is its swiftness:
Once the algorithm has stabilized, how fast does it
fire given that some process receives a $\go$~input?
%yoram2
In addition to solving the self-stabilizing firing squad problem,
the algorithm presented in this paper
%The algorithm presented in this paper
%not only solves the self-stabilizing firing
%squad problem, it
is also optimal in terms of both its stabilization
time, and its swiftness.

\noindent
The main contributions of this paper are:
\begin{itemize}
\item A self-stabilizing variant of the firing squad problem is
  defined, and an algorithm solving it in the case of crash failures
  is given.
\item The proposed algorithm, called \fireAlg, is shown to be optimal
  both in terms of the time it requires to stabilize and  in terms of
  the time it takes, after stabilization, to fire in response to
  a \go input.
\item Finally, the optimality is demonstrated in a fairly strong
  sense: For every possible failure pattern, both stabilization time
  and swiftness are the fastest possible, in any correct algorithm.
In extreme cases this enables   stabilization in two rounds and firing
in one round.
  \end{itemize}

The rest of the paper is organized as follows.
\sectionref{sec:model} describes the model and defines
%a simplified version of
the problem at hand. \sectionref{sec:lower} provides lower
bounds for the optimality properties. \sectionref{sec:upper} describes
the proposed solution, \fireAlg, and proves its correctness and
optimality.
%\sectionref{sec:strictsafety}\dd{this section was commented out}
%defines a stricter version of the problem and shows that \fireAlg
%solves it as well.
Finally, \sectionref{sec:discussion} concludes with a discussion.

\section{Model and Problem Definition}\label{sec:model}
The system consists of a set $\p=\{1,\ldots,n\}$ of processes.
Communication is done via message passing, and the network is
synchronous and fully connected. The system starts out at
%yoram2
time\footnote{All references to ``time'' in this paper refer to non-negative
integer times.}
%time\footnote{When discussing any time $k$ throughout this work,
%$k$ is assumed to be a non-negative integer.}
$k=0$, and
a communication round $r$ starts at time $k=r-1$ and ends at time
$k=r$. At time $k$ each
process computes its state according to its state at time $k-1$,
%yoram2 added
the
internal messages it received by time $k$ (sent by other processes at time $k-1$)
and external inputs (if any) that it received at time $k$.
%\dd{should we define it as between k-1 and k?}
%\ezra{it is clearer when defined as a precise time}
In
addition, at any time
$k\geq0$ a process can produce an external output (such as
``firing'').

Let $\I^k_p \in \{0, 1\}$ represent the external input of process $p$ at
time $k$. We say that $p$ received an external $\go$~input at time $k$ if
$\I^k_p=1$; Otherwise, (if $\I^k_p=0$), we say
that $p$ {\em did not} receive a $\go$ input. Let $\I_p =
\{\I^k_p\}_{k=0}^\infty$, let $\I^k =\{\I_p^k\}_{p=1}^n$ and let $\I
=\{\I_p\}_{p=1}^n$. $\I$ is ``the input pattern'', and $\I^k$ is the (joint)
input at time $k$. In a similar manner define $\OO^k_p \in \{0,1\},
\OO_p, \OO^k$ and $\OO$ as the output pattern. If $\OO^k_p=1$ we say
that $p$ fires at time $k$, and
if $\OO^k_p=0$ we say $p$ does not fire at time $k$.
It will be convenient to say that a fire action occurs at time~$k$ if
$\OO_p^k=1$ for some process~$p$, and similarly that a $\go$ input is
received at time~$k$ if $\I_p^k=1$ for some~$p$.

Denote by $t$ an \emph{a priori} bound on the number of faulty processes in
the system.
For ease of exposition, we assume that $t<n-1$, so that there are
at least two processes that need to coordinate their actions.
We assume the {\em crash} failure model, in which a
faulty process $p$ does not send any messages after its failing
round; it behaves correctly before its failing round, and sends
an arbitrary subset of its intended messages during
%yoram2
its
%the
failing round.
\hide{is failed. Formally, at time $k$ each process $p$ is either faulty or
non-faulty. If $p$ is faulty at
time $k$ then $p$ is faulty at all times $\geq k$, and $p$ does not
send any messages during all rounds $\geq k+1$.\dd{again round and
  time}
\ym{But there is no problem with them here.}}

A failure pattern describes for each time $k$ which processes
have failed by time~$k$, and for each process that fails in round~$k$
(\ie did not fail by time~$k-1$), which of its outgoing communication
channels are blocked (and hence do not deliver its messages) in round~$k$.
Notice that a process may fail in round $k$ even if all of its messages
are delivered.
We denote a failure pattern by~$\F$, and by $\F^k$ the set of
processes that fail in~$\F$ by time~$k$.
%ezra1: : -> ;
Observe that $\F^k \subseteq \F^{k+1}$; in the crash failure model
failed processes do not recover. Similarly, we use $\G^{k}=\p \setminus \F^k$
to denote the set of processes that are non-faulty at time $k$.
Finally, $\G$ will denote the set of processes that remain non-faulty
throughout~$\F$, \ie
$\G = \bigcap_{k=0}^\infty\G^k$. Notice that the set~$\G$ is always
defined in terms of a failure pattern~$\F$, which is typically clear
from the context.

In addition to crashes, there are also transient
%yoram2 This repeats in the coming sentences
faults.
%faults, meaning that the system may start at an arbitrary state.
Formally, we denote by $\s_p^k$ the
state of a process $p$ at time $k$. We denote by $\s^k=(\s_1^k,\dots,
\s_p^k, \dots, \s_n^k)$ the state of the entire system
at time $k$. Transient faults are captured by the assumption that the
system may start from any (arbitrary) state, and there is some round
$r$ such that for all rounds $r' \geq r$ the intended algorithm
operates as written.
%yoram2 UNCLEAR I didn't understand the next sentence.
%               In particular, how does $r$ fit here? Please rephrase.
In other words, for any
possible state $S$, if $\s^0=S$ then eventually (starting from some round $r$) the
algorithm operates correctly.

For the following analysis, each algorithm $\A$ is assumed to have an
initial state $\s_{init}^\A$. For self-stabilizing algorithms, we fix an
arbitrary state as $\s_{init}^\A$ (as the algorithm should converge
starting from any initial state).
The \emph{a priori} bound of~$t$ on the number of failures is
assumed to be hard-wired into the algorithm, and is not affected by
transient faults. Such an algorithm is assumed to be executed only
in the context of failure patterns in which at most~$t$ processes
crash. For such failure patterns~$\F$, the algorithm $\A$ produces an
output pattern $\OO$ starting from state
$\s$ given an input $\I$; we denote this
output pattern by $\OO = \A(\s, \I, \F)$.

Informally, the Firing Squad problem requires that: (1) all processes
fire together (\simul); (2) if a $\go$ input is received then a fire
action occurs (\liveness); and (3) the number of fire actions is not
larger than the number of received $\go$ inputs (\safety). Formally,
%\dd{the 3rd differ from the one we have in the intro which is actually the 2nd here - i do see the resemblance between the informal and the 3rd item.  you  may want to add that since we do not quantify the time to react, the 3rd needs to take this form.}\ezra{I agree that it is clear}
%\ezra{Three requirements have been presented informally and formally in Definition 1. While the informal presentation of the first two requirementsare is clearly reflected in Definition 1, the third requirement, safely, is not clear as item 3 of Definition 1.}

\begin{definition}\label{def:prop}
Let $\OO = \A(\s, \I, \F)$ and let $\G$ denote the set of
%yoram2 TYPO
processes
%process
that remain non-faulty throughout $\F$. We say that~$\OO$ satisfies the $\FS(k)$
properties (capturing correct firing-squad behavior from time~$k$ on)
w.r.t.~$\I$, $\F$, and  $\OO$, if the following conditions hold for
all $k' \geq k$:
\begin{enumerate}
\item (simultaneity) If $\OO_p^{k'}=1$ for some $p\in\p$ then
$\OO_q^{k'}=1$ for all ${q\in\G}$;
%$\forall_{q\in\G^{k'}}\OO_q^{k'}=1$;
\item (liveness) If $\I_p^{k'}=1$ for some $p\in\G$, then there is $k''>k'$ s.t.
 $\OO_p^{k''}=1$;
\item (safety) The number of times $k''$ satisfying
$k\le\hk\le k'$ at which a fire action occurs at~$k''$
is not larger than the number of times $h$ in the range $0\le h< k'$ at which
$\go$ inputs are received.
\end{enumerate}
\end{definition}

We can use the $\FS(k)$ properties to define when an algorithm solves
the firing squad problem in a self stabilizing manner. We first use it
to define the stabilization time of an algorithm as follows:

\begin{definition}[Stabilization time]
\label{stabilizationtime}
The {\em stabilization time} of $\A$ on $\s$, $\I$ and $\F$,
denoted by $\stab(\A,\s,\I,\F)$,
is the minimal~$k\ge 0$ such that $\FS(k)$ holds with respect to
$\I$, $\F$, and $\OO = \A(\s, \I, \F)$. (If $\FS(k)$
holds for no finite~$k$, then $\stab(\A,\s,\I,\F)=\infty$.)
\end{definition}

Notice that the \safety property in $\FS(k)$ relates outputs
starting from time~$k$ to inputs starting from time $0$.
%ezra1: . -> :
Here's why:
%and for outputs starting from time $k$.
Since we consider time $0$ to be
the point at which transient errors end, if the system starts in a state
in which ``it appears as if'' \go inputs were received before time~0,
the good processes may fire
after time~0 without a \go message actually having been received.
%a few times and violate \safety of $\FS(0)$.
Once all firings induced by such
``phantom'' \go  inputs have
occurred, we can legitimately require firing events to happen only
in response to genuine \go message receipts.
We thus think of the stabilization time, at which in particular
 the safety property of $\FS(k)$ holds, as one after which no firing
 will occur in response to phantom \go messages.
Rather, every firing will be justifiable as a response to some \go
message received at or after time~0.

%ezra: changed the following def a bit
\begin{definition}[SSFS Algorithm]\label{def:SSFS}
An algorithm $\A$ solves the Self stabilizing Firing Squad problem
($\A$ is an SSFS algorithm, for short)
if there exists a $k < \infty$ such that $\stab(\A,\s,\I,\F) \leq k$ for
every system state $\s$, input
pattern $\I$ and failure pattern $\F$.
\end{definition}

Observe that in a setting with no transient faults, an algorithm $\A$
solves the (non-self-stabilizing) Firing Squad  problem if it
satisfies $\FS(0)$ with respect to ~$\I$, $\F$, and  $\OO$, for every
$\I$, $\F$ and $\OO=\A(\s_{init}^\A,\I,\F)$.

\hide{\begin{remark}\label{remark:safety}
  The \safety property in \definitionref{def:prop} states that if some
  process $p$ fires then beforehand some process $p'$ received a $\go$
  input. Thus, the \safety requirement is quite relaxed as once $p'$
  received a $\go$ input, any process may fire whenever it wishes.
However, when considering the self-stabilizing version of the FS
problem the \safety requirement acquires a stricter sense. Suppose
that an algorithm fires forever once there are no more external $\go$
inputs. Therefore, there is a memory state $\s$ that with input
containing no $\go$s will produce an output pattern $\OO$ that for
every $k$ there is some $k' \geq k$ such that $\OO$ has a fire at time
$k'$. Thus, the algorithm does not solve the SSFS problem.
  The above explanation shows that if an algorithm solves the SSFS
  problem, then it actually must adhere to a stronger property than
  \safety. We use \safety as defined to simplify the proofs and
  discussion. In \sectionref{sec:strictsafety}\dd{}need to take care a
  stricter version of the \safety property is defined and discussed
  (the proposed solution, \fireAlg, adheres to both versions of
  \safety).
\end{remark}
}

Notice that \definitionref{def:SSFS} implies that any SSFS algorithm $\A$
has at least one memory state from which
the firing squad properties are guaranteed to hold.
%(as long as there are  no additional transient faults) all of FS's
%properties hold.
Denote one of these memory states by $\s_{stab}^\A$,
or simply $\s_{stab}$ when $\A$ is clear from the context.

\subsection{Optimality Measures}
In this work we are interested in finding an optimal SSFS algorithm.
We start by defining stabilization time optimality, which
measures how quickly algorithm $\A$ stabilizes.

%\begin{definition}
%Let $\A$ be an algorithm that solves the SSFS problem, and let
%$\psi \in \{\mbox{simultaneity, liveness, safety}\}$.
%We denote by \Time($\A, \s, \I, \F, \psi$) the minimal
%time $k$ such that $\psi$ holds w.r.t. $\I$, $\F$, and $\OO=\A(\s, \I, \F)$
%starting from time $k$ on.
%\end{definition}

\begin{definition}\label{def:AlessA}
An SSFS algorithm $\A$ is said to {\em optimally stabilize}
if the following holds for every SSFS algorithm $\BB$ and
every failure pattern~$\F$:
\[\max_{\s,\I}\{\stab(\A,\s,\I,\F)\} ~~\leq~~
\max_{\s,\I}\{\stab(\BB,\s,\I,\F)\}\;.\]
%ezra1: removed
%\ezra{Explaining in more details the intuition beyond the choice of that optimality definition could be nice. I wonder how the results of the paper change if one take stronger/weaker definitions of optimality (f.i. considering $max_{i,f,s}$ or having the inequality for any i,f,s).}
\end{definition}

\definitionref{def:AlessA} defines optimality of an algorithm $\A$
with respect to its stabilization time, \ie how quickly $\A$
starts to operate according to all of the $\FS$ requirements.
%ezra1: added following pgrp
The intuition
behind defining optimality in terms of worst-case $\s$ and $\I$ is to avoid
algorithms that are ``specific'' to an initial memory state or input pattern.
Thus, by requiring optimality in the worst-case we ensure that the algorithm
cannot be hand-tailored to a specific setting, but rather needs to solve the SSFS
problem in a ``generic'' manner.

%ezra1: removed
%The following definition deals with how quickly $\A$ responds to a
%$\go$~input, once the algorithm has stabilized.
%given that  the alall the $\FS$ requirements hold.

We now turn to the issue of comparing the responsiveness of distinct
firing squad algorithms. Specifically, we are concerned with how
quickly an algorithm fires after a \go message is received
%ezra1: added
(once the algorithm has stabilized).
For simplicity, we consider receipts of \go by non-faulty processes, since
the problem specification forces a firing following such a
receipt. Another subtle issue is that if \go messages are received in
different rounds between which there is no firing, then it may be
difficult to figure out which \go message the next firing is
responding to. Again for simplicity, we will be interested in what
will be called {\em sequential} input patterns, in which a \go is not
received before all previous \go's have been followed by firings.
More formally, we define:
\hide{exactly one \go is received between any they must generate a firing
To measure how quickly an algorithm $\A$ fires, specific input
patterns are used.
To illustrate the reason for that consider
an input pattern that has an external $\go$ input at every time unit: it will be hard
to define and measure how quickly $\A$ fires given an external $\go$ input. Therefore, responsiveness is measured only with respect to  {\em sequential}
input patterns, as defined below.}

% The {\em stabilization time} in a run of an SSFS algorithm $\A$
% on $\s$, $\I$ and $\F$,
% denoted by $\stab(\A,\s,\I,\F)$,
% is the minimal~$k\ge 0$ for which the Firing Squad
% properties hold from time~$k$ on.

\begin{definition}[Sequential inputs]
Let~$\A$ be an SSFS algorithm.
We say that the  input $\I$ is sequential with respect to
($\A$, $\s$, $\F$) if
(i) no $\go$ inputs are received according to~$\I$
at times $k<\stab(\A,\s,\I,\F)$,
(ii) $\go$ inputs are received in $\I$ only by processes from~$\G$,
and
(iii) if $k_1<k_2$ and $\go$ inputs are received at both~$k_1$
and~$k_2$,
then there is an intermediate time $k_1<k'\le k_2$ at which a fire action
occurs.
\end{definition}

The following definition formally captures the number of firing events
that occur between the stabilization time and a given time~$k$.

\begin{definition}
  Let~$\A$ be an SSFS algorithm and let $\OO\!=\!\A(\s, \I,\F)$.
We define $\#[(\A,\s, \I,\F),k]$ to be the number of rounds~$k'$
in the range %satisfying that
 $\stab(\A,\s,\I,\F)\le k'\le k$ such that %and that
$\OO^{k'}_p=1$ holds for some process~$p$ (\ie a firing
occurs at time~$k'$).
\end{definition}

By definition, if $k < \stab(\A,\s,\I,\F)$ then $\#[(\A,\s, \I,\F),k]=0$.
With the last two definitions, we are now able to formally
%We can now use the notion of sequential inputs to
compare the
responsiveness of different SSFS algorithms:

\begin{definition}[Swiftness]\label{def:quickness}
Let~$\A$ and $\BB$ be SSFS algorithms. We say that $\A$ is {\em at
  least as swift as}~$\BB$ if
%denoted $\A \leq_{\quick}\BB$, if
$\A$ fires at least as quickly as~$\BB$ on all sequential inputs.
Formally, we require that for every failure
pattern~$\F$, input~$\I$, and states~$\s_\A$ of~$\A$ and $\s_\BB$
of~$\BB$, the following holds. If $\I$ is sequential
both with respect to ($\A$, $\s_\A$, $\F$) and with respect to
($\BB$, $\s_\BB$, $\F$), then
$\#[(\A,\s_\A,\I,\F),k]\ge \#[(\BB,\s_\BB,\I,\F),k]$
holds for every time~$k$.
An SSFS algorithm $\A$ is optimally swift if it is
at least as swift as $\BB$ for every SSFS algorithm~$\BB$.

\end{definition}

%Given Now that both optimality measures are defined
%(\definitionref{def:AlessA}, \definitionref{def:quickness})
We are now in a position to state the main result of the paper:
The \fireAlg algorithm of~\figureref{figure:fireAlg}
is an SSFS algorithm (\theoremref{thm:ssfs}),
is optimally stabilizing (\theoremref{thm:stabotimal})
and is optimally swift (\theoremref{thm:quickoptimal}).

\section{Lower Bounds}\label{sec:lower}
In this section we provide lower bounds for the stabilization time
and for the swiftness of any SSFS algorithm $\A$. The lower bounds
build upon previous results in the field of simultaneous agreement.

Recall that if $\A$ is a non-self-stabilizing Firing Squad algorithm,
then $\stab(\A, \s_{init}^\A, \I, \F) = 0$ for all $\I$ and
$\F$. Therefore, in the non-self-stabilizing case,
it only makes sense to compare algorithms in terms of their ``swiftness.''
In a non-self-stabilizing setting, the firing squad
protocol $\ccfs$ (based on $\concon$ \cite{MM}) is optimally swift.
We will use it as a benchmark and yardstick for expressing and
analyzing the performance of self-stabilizing firing squad protocols.
To compare the performance of different algorithms,
we make use of the following definitions.
%
% encapsulating the limits a failure
%pattern $\F$ induces on the \safety of a SSFS algorithm.

\begin{definition}\label{def:x}
%  Given a failure pattern $\F$.
  We denote by $\x(\F, k)$ the number of processes known at time $k$
  to be faulty by the processes in $\G^k$
  in a run of $\ccfs$  with failure pattern~$\F$.
\end{definition}
Intuitively, $\x(F,k)$ stands for the number of failures that are
{\em   discovered} by time~$k$ in a run with pattern~$\F$.
We remark that $\x(\F,k)$ is well-defined, because the same number of
faulty processes are discovered (at the same times) in all runs
of~$\ccfs$ that have
failure pattern~$\F$. Moreover, since $\ccfs$ detects failures as a
full-information protocol does, no algorithm $\A$ can discover more
failed processes than $\ccfs$ does (see~\cite{DM}). Thus, $\x(\F,k)$
is an upper bound on the number of failed process discovered by time
$k$ by any algorithm $\A$.

$\ccfs$ makes essential use of a notion of {\em horizon}, which
is roughly the time by which past events are
guaranteed to become common knowledge. This motivates the following
definitions.

\begin{definition}[Horizons]
  Given a failure pattern $\F$, the {\em horizon distance} at time
  $k$, denoted by $\rh(\F, k)$, is $t+1-\x(\F, k)$.
  The {\em absolute horizon} at time $k$, denoted $\ah(\F, k)$, is
  $k+\rh(\F, k)$.
\end{definition}

While the absolute horizon is an upper bound on when events become
common knowledge, the publication time is a lower bound
on this time. It is defined as follows:

\begin{definition}[Publication Time]
  Given a failure pattern $\F$, the {\em publication time} for
  (time)~$k$, denoted by $\bb(\F, k)$, is $\min_{k'\geq k} \{\ah(\F,
  k')\}$.
\end{definition}

When $\F$ is clear from the context, it will be omitted from $\x(k)$,
$\rh(k)$, $\ah(k)$ and $\bb(h)$.

As shown in \cite{MM}, for a given failure pattern $\F$, a $\go$
input received at time $k$ is ``common knowledge'' not before time
$\bb(\F, k)$. Thus, for a specific algorithm $\A$, the publication time
for~0
%border at time $0$
bounds (from below) the time $k$ at which the first firing action can
occur in $\OO=\A(\s_{stab}, \I, \F)$.

The publication time $\bb(\F, k)$ is a generalization of
notions developed in \cite{DM} for Simultaneous (single-shot,
non-stabilizing) Consensus.
In that paper, a notion of the {\em waste} of $\F$ is defined,
and information about initial values---which
can be viewed in our setting as being  about external inputs
at time~0---becomes common knowledge at time $t+1-\textit{waste}$. In
our terminology, this occurs precisely at the publication time
$\bb(\F,0)$ for events of time~0.

The intuition behind the first lower bound is that if $\ccfs$ receives
a $\go$ input at time $0$, then it fires at time $\bb(0)$ (\lemmaref{lemma:ccfs0}).
Since $\ccfs$ is optimal, an SSFS algorithm $\A$ cannot fire faster. Therefore, if we consider
$\A$ starting in a memory state where $\A$ ``thinks'' it received a $\go$ input
$1$ round ago, $\A$ will fire not before time $\bb(0)-1$.
The formal proof
appears in the proof of \theoremref{theorem:lowerbound1}.

\begin{lemma}\label{lemma:ccfs0}
  Let $\F$ be any failure pattern and let $\I$ be an input pattern for
  which $\I_q^k=0$ for every process $q$ and time $k \geq 0$, except
  for one process~ $p \in \G$ for which  $\I_p^0=1$.
%and ``1'' otherwise.
The first   fire action of $\OO=\ccfs(\s^{\ccfs}_{init}, \I, \F)$
occurs at time $\bb(\F, 0)$.
\end{lemma}
\begin{proof}
  A result of the work done in \cite{MM}.
\end{proof}

\begin{notation}
For input $\I$ and an integer $i \geq 0$, denote by $\I(i
\rightarrow)$
the input pattern that is obtained by excluding the first~$i$ rounds
of~$\I$.
Formally, $\I(i \rightarrow)^k = \I^{k+i}$ for all $k\ge 0$.
Similarly denote $\F(i\rightarrow)$  (w.r.t.~$\F$).
\end{notation}

\begin{lemma}\label{lemma:filurecompare}
Let $\F$ be a failure pattern. Let $\F'$ be a failure pattern with no
faults at time $k=0$
and $\F'(1\rightarrow)=\F$. Then $\bb(\F',0) \geq \bb(\F, 0)$.
\end{lemma}

\begin{proof}
For every time $k$ we have  that $\x(\F', k) \leq \x(\F,
k)$. Therefore,  $\ah(\F', k) \ge \ah(\F, k)$
holds for all~$k\ge 0$. Thus, $\min_{k\geq 0} \{\ah(\F', k)\} \geq \min_{k\geq 0}
\{\ah(\F, k)\}$, \ie $\bb(\F',0) \geq \bb(\F, 0)$.
\qed\end{proof}

Following is the first lower bound result, stating that
the worst case stabilization time of every
SSFS algorithm $\A$ is at least $\bb(0)$.

\begin{theorem}\label{theorem:lowerbound1}
$\max_{\s, \I}\{\stab(\A,\s,\I,\F)\} \geq \bb(\F, 0)$
holds for every SSFS algorithm $\A$
%that solves the SSFS problem
and every failure pattern $\F$.
\end{theorem}
\begin{proof}
To prove this theorem, we find a state $\s$ and input $\I$ such that $\stab(\A,\s,\I,\F) \geq \bb(\F, 0)$.
Since $\A$ solves the SSFS problem there is a memory state $\s_{stab}$
from which all of the $\FS$ properties hold.

Let $p\in\G$ be a process that is non-faulty throughout $\F$, and consider the following input path $\I'$: for all $q, k$ it holds that ${\I'}^k_q =0$ except for ${\I'}_p^0=1$.  Consider $\F'$ to be a failure pattern with no failures at time $k=0$ (\ie $\F'^0=\emptyset$) and $\F'(1\rightarrow)=\F$ for the rest. Due to \textit{``liveness''}, $\A$'s run from $\s_{stab}$ with input $\I'$ and failures $\F'$ will eventually fire; denote the firing time as $k$ (\ie $\OO^{k}_p=1$ for some process $p$).

By \lemmaref{lemma:ccfs0}, $\bb(\F', 0)$ is the optimal time for simultaneous firing, and since starting from $\s_{stab}$ all properties hold, including \simul, it holds that $k \geq \bb(\F', 0)$.

Consider memory state $\s$ of $\A$ after executing a single round with
$\I'$ as input and $\F'$ as failure pattern and $\s_{init}$ as
starting memory state. Consider the run of $\A$ from $\s$ with input
$\I$ and failure pattern $\F$. $\A$ must fire at time $k-1$,
as it cannot distinguish the run from $\s_{init},\I',\F'$ and from
$\s,\I,\F$ . By \lemmaref{lemma:filurecompare}, $\bb(\F',0) \geq \bb(\F, 0)$, and
therefore $\A$ will not fire before time $k-1 \geq \bb(\F',0)-1 \geq
\bb(\F,0)-1$. However, notice that $\I$ contains only ``0''
inputs, implying that \safety does not hold for $\A$ when starting
from $\s$ with input $\I$ and failure $\F$ for the first $\bb(\F,0)-1$
rounds. \Ie \safety can hold starting from time $\bb(\F,0)$ and
on. Therefore, $\max_{\s, \I}\{\stab(\A,\s,\I,\F)\} \geq \bb(\F, 0)$.
\qed\end{proof}

Our second lower bound result, informally stating that
any SSFS algorithm cannot fire faster than $\ccfs$, is captured by
the following theorem. (Notice that the claim is made with respect to
sequential input patterns.)
\begin{theorem}\label{theorem:lowerbound2}
  Let $\A$ be an SSFS algorithm, $\I$ a sequential input, $\F$ a
  failure pattern and
  $\OO=\A(\s_{stab}, \I, \F)$. For every $k \geq 0$ for which a $\go$ input is received in $\I^k$
  there is no fire action in $\OO$ during times $k'$ satisfying $k < k' < \bb(\F, k)$.
\end{theorem}
\begin{proof}
  Suppose by way of contradiction there is such a time $k'$, and
  consider the earliest such time $k'$
  satisfying $k < k' < \bb(\F, k)$ for which a fire action occurs in
  $\OO^{k'}$. Denote by $\s^k$ the memory state of $\A$ at time $k$.

  Since $\A$ started to run from $\s_{stab}$, $\FS(0)$ holds with
  respect to $\I,\F$ and $\OO$.
  Since $\I$ is sequential, and $k'$ is the minimal time for which
  $\OO$ has a fire action after time $k$, we have that $\I(k
  \rightarrow)$ contains a $\go$ input at time $0$ and does not
  contain a $\go$ input until time $k'$. Therefore,
  $\OO=\A(\s^k, \I(k \rightarrow), \F(k \rightarrow))$ will have its
  first fire action at time $k'-k$.

  From $\ccfs$'s optimality and together with \lemmaref{lemma:ccfs0},
  $\A$ cannot fire before time $\bb(\F(k \rightarrow),0)$.
  Thus $k'-k \geq \bb(\F(k \rightarrow),0)$ leading to $k' \geq k+ \bb(\F(k \rightarrow),0)$.
  By definition of $\bb$ and $\F(k \rightarrow)$ we have that
  $\bb(\F, k) \leq \bb(\F(k \rightarrow),0)$, contradicting the assumption that $k < k' < \bb(\F, k)$.
\qed\end{proof}

%\begin{lemma}
%For every algorithm $\A$ that solves the SSFS problem and for every failure pattern $\F$, it holds that $$\max_{\s, \I}\{\Time(\A,\s,\I,\F,\simul)\} \geq 1\;.$$
%\end{lemma}
%\begin{proof}
%Since $\A$ solves the SSFS problem there is a memory state $\s'$ and
%an input pattern $\I$
%for which $\A(\s',\I,\F)$ a fire action occurs at time $0$. Similarly, there is a memory state $\s''$
%for which no fire action occurs at time $0$ in $\A(\s',\I\,\F)$.
%
%Consider the state $\s$ constructed in the following way: for some $p \in \G$ let $\s_p={\s'}_p$ and for the other processes $q$ let $\s_q={\s''}_q$. Clearly, in $\A(\s,\I\,\F)$ some process fire at time $0$ and some process does not. Thus, \simul does not hold at time 0.
%\qed\end{proof}
%
%\ezra{What can we say about \simul and \liveness? any interesting lower bounds?}

\section{Solving SSFS}\label{sec:upper}
The algorithm \fireAlg in \figureref{figure:fireAlg}
is an SSFS algorithm that is both optimally stabilizing
and is optimally swift. For swiftness, the algorithm is
based on the approach used in the $\ccfs$ algorithm, in
which the horizon is computed by monitoring the number of
failures that occur, and a firing action takes place when
the receipt of a \go becomes common knowledge.
The horizon computation at a process~$p$
makes use of reports that~$p$ receives from other processes
regarding failures that they have observed.
Following a transient fault, the state of a process may contain
arbitrary (including false) information about failures. In the crash
failure model, a process~$q$ will learn about (truly) crashed processes
in the first round. Consequently, $p$ will compute a correct horizon
one round later, once it receives reports from all such processes.
Roughly speaking, this can be used as a basis for a (nontrivial)
solution that stabilizes within two rounds of the optimal time.

In order to improve on the above and obtain an optimal algorithm, \fireAlg
employs a couple of subtle consistency checks.
The first one involves checking the information obtained from other
processes regarding failures they observed before the current round
started. In the crash failure model, every failure observed by~$q$
by time~$k-1$ must be directly observable by~$p$ no later than
time~$k$. So if the set of failures reported to~$p$ contains failures
that~$p$ has not directly observed, then it must be
time~$k\le 1$, and~$p$ will use the set of failures that it has
directly observed in computing the horizon, instead of the set of
reported failures.
A subtle proof shows that, in this case, the computed horizon
works correctly if $k=1$, which is crucial for the algorithm's
stabilization optimality.
The second consistency check is based on the fact that in
normal operation the horizon distance is (weakly) monotone
decreasing. The local state contains information about previous
horizon computations, and our second consistency check forces
it to satisfy weak monotonicity.

\begin{figure*}[t]\center

\begin{minipage}{4.8in}
\hrule \hrule \vspace{1.7mm} \footnotesize
%\begin{alltt}
\setlength{\baselineskip}{3.9mm} \noindent Algorithm \fireAlg$\!(t)$
 \vspace{1mm} \hrule \hrule
\vspace{1mm}

\begin{tabular}{ r l }
\linenumber & {\bf do} forever:\hfill\textit{/* executed on process
  $p$ at time $k$ */}\\
& \hfill\textit{/* process $p$ is unaware of the value of $k$ */}\\

\linenumber & \tb {\bf receive} all available ($\req_q, \fail_q,
\view_q$) messages from process $q\in\p$;\\
\\

& \tb \textit{/* update variables according to messages of round $k$ and external input */}\\
\linenumber & \tb {\bf set}  \req$[0] := \I_p^{k}$;  \\
\linenumber & \tb {\bf for} $1 \leq i \leq t+1$: {\bf set} \req[i] :=
$\max_q \{\req_q[i-1]\}$; \\
\linenumber & \tb {\bf set} $\failp := \bigcup_q \fail_q$; \\
\linenumber & \tb {\bf set} \fail \ := all processes that $p$ did not
hear from this round;\\
\linenumber & \tb {\bf for} $1 \leq i \leq t$: {\bf set} \view$[i-1]$ := $\min_q\{\view_q[i]\}+1$; \\

\\
& \tb \textit{/* calculate horizon at time $k-1$ */}\\
%ezra1: added consistency check I
\linenumber & \tb {\bf set} \horz\ := $t+1-\min\{|\failp|, |\fail|\}$; \hfill\textit{/* consistency check I */}\\
\linenumber & \tb {\bf set} \view[\horz-1] := 1; \\
%ezra1: added check II
\linenumber & \tb {\bf for} $0 \leq i  \leq t$: {\bf set} \view$[i]$
:= $\max\{\view[i],\horz-i\}$; \hfill\textit{/* check II */}\\

\\
& \tb \textit{/* should we fire? */}\\
\linenumber & \tb {\bf if} for some $i' \geq \view[0]$ it holds that
$\req[i']=1$ then  \\
\linenumber & \due {\bf for} $i' \leq i'' \leq t+1$: {\bf set}
$\req[i''] := 0$; \\
\linenumber & \due {\bf do} ``Fire''; \\
\linenumber & \tb {\bf fi}; \\
\\
& \tb \textit{/* send round $k+1$ messages to all processes  */}
\\
\linenumber & \tb {\bf send} (\req, \fail, \view) to all; \\
\linenumber & {\bf od}.

\end{tabular}
\hrule
\vspace{1mm}
\tb {\bf Clean up:}\\
\tb $\req$ contains only $\{0,1\}$ values.
$\view$ contains only values $\in \{0, \dots, t+1\}$.
\normalsize \vspace{1mm} \hrule\hrule
\end{minipage}

 \caption{\fireAlg: a self-stabilizing firing squad
   algorithm.}\label{figure:fireAlg}
\end{figure*}

We now turn to describe the details of $\fireAlg$.
The following discussion and
lemmas are stated w.r.t.\ the algorithm and its components.
For a variable {\it var}, we denote by $\mbox{\it var}_p^k$ the value of
{\it var} at process $p$ after the computation step at time $k$.

Each process $p$ has a vector $\req_p[i]$, which represents
$p$'s information about a $\go$~input received by some process $i$
time units
ago; and this request was not fulfilled yet. More precisely, if
$\req_p^k[i]=1$, then
some process received a $\go$~input  at time $k-i$, and
no firing action occurred
between time $k-i+1$ and time $k$. The vector $\req$ contains values for the
previous $t+1$ time units and the current time; a total of $t+2$
entries.

In addition, each process has a set $\fail$, which consists of the processes
it has seen to be failed in the current round. That is, at time $k$,
process $p$'s $\fail_p^k$ set contains all processes that process $p$
did not received messages from during round $k$ (\ie messages sent at time
$k-1$). $\failp$ is the union of all
$\fail$ sets (as received from other processes) of the previous
round. That is, at time $k$, ${\failp}_p^k$ is the union of
$\fail_q^{k-1}$ as computed at time $k-1$ by every process $q$ that $p$
received messages from during round $k$.

Finally, each process keeps track of a vector $\view$. If
$\view_p^k[i]=z$ it means that at time $k+i$, data from time $k-z$ is
common knowledge. The vector $\view$ contains $t+1$ entries, for the
current round and the coming $t$ rounds.

\hide{
\subsection{proof outline}
\begin{itemize}
    \item If all processes have the same values of \req, \fail, \view\ then the algorithm should work due to previous proofs.

    \item once there is a clean round, all processes have the same values of \req, \fail and \view.

    \item finally, to show optimality, we need to show that if there
      isn't a clean round, then non of the properties hold.
\end{itemize}
} %hide

For ease of exposition every process $p$ is assumed to send
messages to itself. Moreover, a process executing the algorithm is
unaware of the current round number.
We refer to such rounds using numbers~$k$ etc. for ease of
exposition in describing and analyzing the algorithm.

\subsection{Correctness Proof}
A central notion in the analysis of simultaneous actions under
crash failures is that of a {\em clean round} \cite{DM}.
In the non-stabilizing setting, a round~$r$ is clean according to failure
pattern~$\F$ if no process considered non-faulty by all
processes at time $r-1$ is known to be faulty by one or
more (non-crashed) processes at time~$r$.
In a setting that allows transient faults, we use a slightly different
definition for the exact same notion.
Consider a process~$p$ that fails in round~$k$. We say that $p$ fails
{\em silently} in round~$k$ if it is not blocked according to~$\F$
from sending messages in round~$k$ to any of the processes
$q\in\G^k$. Thus, no process surviving round~$k$ can detect~$p$'s
failure in this round.
\begin{definition}[Clean Round]
Round $r$ in failure pattern $\F$ is a {\em clean round} if
(i) no process fails silently in round~$r-1$, and
(ii) all processes (if any) that fail in round~$r$ fail silently.
\end{definition}
%yoram
This definition of a round $r$ being clean in $\F$ coincides with
the standard definition of clean rounds previously used in
%A round~$r$ is considered clean in~$\F$ according to this definition
%iff it is clean according to the standard definition for
non-stabilizing systems \cite{DM}.
In protocols such as \fireAlg, with the property that
every process sends the same message to all other processes in every
round, all (non-crashed) processes receive the same set of messages in
a clean round (see \lemmaref{lem:allsame}).

We start with an overview of the proof, following a detailed proof. 

\hide{
\subsubsection{Proof overview:}
First, notice that once a clean round has occurred, all processes receive
the same set of messages, and different processes agree on the value of $\req$ (except for $\req[0]$).
Moreover, in the following round all processes agree on
%yoram cut
%on
the value of $\req$ perhaps except for
the value of $\req[0]$ and $\req[1]$. In a similar manner, $k$ rounds after a clean round
the values of $\req[k+1],\req[k+2],\dots$ are the same at all processes.

Second, consider the value of $\view^k[0]$. By \lineref{6},
$\view^k[0]$ equals the value of $\view^{k-1}[1]+1$.
In a similar manner,
%yoram
if
%suppose that
$\view^k[i]$ is updated
%yoram
by
%due to
\lineref{8}
then $\view^{k+i}[0]=\view^k[i]+i$.
%yoram
If
%Suppose
$k$ was the last clean round prior to round $k+i$,
%yoram
then $\view[0] = i+1$ holds at time $k+i$.
%thus by round $k+i$, the value of $\view[0] = i+1$.
Together with the claim from the
previous paragraph, we have that once there was a clean round, if
different processes agree on the value of $\view[0]$, then they
all agree on the
%yoram
values
%value
%ezra1: \dots -> \cdots
of $\req[\view[0]],\req[\view[0]+1],\cdots$.
%yoram combined paragraphs
Thus, if processes agree on the value of $\view[0]$
%yoram added
then
they are guaranteed to act simultaneously, either firing together
or, together, refraining from firing.
Therefore, we turn our attention to analyzing the behavior of $\view[0]$ at the different
processes.

Intuitively, the reason the above discussion does not show that all processes agree on the
%ezra1: added ,
value of $\view[0]$, is
%yoram
the following:
%due to the following fact: e
Even though all processes update
the value of $\view$ in a similar manner (\lineref{6}) each process $p$ updates its own $\view_p$
according to the failures
%yoram added
that
%ezra1: processors -> processes
$p$ has seen in the current round. To show that all processes
have the same value of $\view[0]$ (for all rounds following a clean round) we show two things:
(1) if $\view^k[0]$ is updated in round $k$, then $\horz=1$, \ie $|\failp|=t$. This will be
%yoram
observed by
%true for
all processes, and so they will all set $\view^k[0]=1$;
(2) if $\view^k[0]$ was not
%yoram typo
updated
%updates
in  round $k$, then
%yoram
let $k-i$ be the latest round in which the value of $\view^{k-i}[\horz^{k-i}-1]$ was updated.
%consider the latest round $k-i$ at which the value of $\view^{k-i}[\horz^{k-i}-1]$ was updated.
The proof shows that there must be a clean round between round $k-i$ and round $k$, thus ensuring that all processes
%yoram added
will
agree on the value in $\view[0]$ by round $k$.

Up till now, we have given an overview of the proof that \fireAlg solves the SSFS problem.
To show that it optimally stabilizes and is optimally swift a precise analysis of the
convergence of SSFS is required, along with a proof that SSFS will fire no later than
any other algorithm (on sequential inputs). To illustrate the tools used in those proofs,
we define the following:
%\newpage
\begin{definition}
%\vspace{-1em}
Let

\begin{itemize}
\item $\minHG(\F,k)=\min_{p\in\G}\horz_p^{k}$, and
\item $\ZH(\F, k) = \min_{k' \geq k} \{k'+ \minHG(\F,k'+1) \}$.
\end{itemize}
We write $\ZH(k)$ when $\F$ is clear from the context.

\end{definition}

The main point behind this definition is that $\ZH$ is
the equivalent of $\bb$ with respect to $\fireAlg$ (recall that $\bb$
is computed according to $\ccfs$). In the non-self-stabilizing model $\bb$ is
shown to be a lower bound on when a \go input becomes ``common knowledge''.
Thus, the following two lemmas conclude that \fireAlg is optimally swift.

\begin{lemma}\label{lemma:zklebk}
  $\ZH(k) \leq \bb(k)$, for every $k \geq 0$.
\end{lemma}

\begin{lemma}\label{lemma:firespeed}
  Let input $\I$ be sequential with respect to $(\fireAlg, \s, \F)$.
  If $\I_p^k=1$ for process $p$ at time $k$ then $\OO_{p}^{k'}=1$ for $k<k' \leq \ZH(k)$.
\end{lemma}

Finally, we wish to point out a main difference between the proofs
of the lower and upper bounds in the self-stabilizing model as opposed to the classical model
(with respect to the firing squad problem): in the first round of \fireAlg the value of $\failp$ (the
set of processes
%yoram added
that have
failed in the previous round) might be contaminated. That is, a process
may start in a state where it thinks
that
some other processes are failed, even though they are correct.
Thus, a major property that is used in the classical proofs cannot be used freely in the self-stabilizing
model's proofs: the monotonicity of crash failures. In the classical model, the perceived set of crashed processes
can only increase, while in the self-stabilizing model it may decrease following the first round.

This explains the purpose of \lineref{7}, which is to perform a consistency check,
comparing the reported $\fail_q$ values (from the previous round) to
failures
%yoram added
that are
directly observed by~$p$ in the current round (stored in $\fail_p$). This comparison
%ezra1: rephrased
together with a delicate treatment in the proofs, ensures the optimality of \fireAlg.
That is, to prove that \fireAlg is optimal up to an additive constant of 1 round is much easier than to prove
that \fireAlg is optimal.
%yoram added
We prove the latter, stronger, property.
} %hide

\subsubsection{Proof outline:}
\begin{enumerate}
  \item Once a clean round has occurred,
    different processes agree on the value of $\req$
    (\lemmaref{lem:allsame} and \lemmaref{lemma:oldreq});
  \item Thus, if processes agree on the value of $\view[0]$ they
are guaranteed to act simultaneously, either firing together
    or, together, refraining from firing (\lemmaref{lemma:firetogether});
%  \item the value of $\horz$ does not increase (see \lemmaref{lemma:dechorz}).
  \item \lemmaref{lemma:zkeq1} and \lemmaref{lemma:foralli} show that $\view[0]$
    is the same at all non-crashed process (once a clean round has occurred);
  \item Points 1, 2 and 3 above lead to \lemmaref{lemma:simul}, stating
    that once a  clean round occurs, \simul holds;
  \item \liveness holds by \lemmaref{lemma:liveness};
  \item \lemmaref{lemma:viewhorz} and \lemmaref{lemma:zklebk} lead to
    \lemmaref{lemma:zkleqk} which states that \safety holds starting from
    round $\bb(0)$. This, according to the lower bounds, is optimal;
  \item \lemmaref{lemma:firespeed} (together with \lemmaref{lemma:zklebk})
    shows that \fireAlg fires by time $\bb(k)$ given a $\go$ input
    at time $k$. The lower bound in \theoremref{theorem:lowerbound2}
    implies that this is optimal;
  \item Finally, \theoremref{thm:ssfs}, \theoremref{thm:stabotimal} and
    \theoremref{thm:quickoptimal} show that \fireAlg is an SSFS algorithm
    that optimally stabilizes and is optimally swift.
\end{enumerate}

\begin{lemma} \label{lem:allsame}
If round~$r$ is clean, then the sets $\fail^r$, $\failp^r$, and
the array $\view^r$ are identical for all non-faulty processes.
\end{lemma}
\begin{proof}
In the \fireAlg algorithm every process sends its $\fail$ set and $\view$
array to all other processes in every round.
If round~$r$ is clean, then all processes receive the same information
about the values of $\fail$ and $\view$ in the system. Thus, the
value of $\view$ computed on \lineref{6}, which depends on the
$\view_q$ values received in the current round, is the   same for all
$p\in\G$. Similarly,  value of $\failp$
  calculated on \lineref{4}, which depends on the
$\fail_q$ sets received is the same at all $p\in\G$.
Finally, in a clean round, all non-faulty processes receive messages
from the same set of processes. As a result,
 the value of \fail\ (computed on
  \lineref{5})\ is the same  all $p\in\G$.
Since changes to \fail, \failp\
  and \view \ performed  on \lineref{7-13} depend only on the values of
  \fail, \failp\ and \view, the same changes are performed by all non-faulty
  processes. The claim follows.
\qed\end{proof}

\begin{lemma}\label{lemma:oldreq}
Let $r$ be a clean round, let $0\le d\le t$ and let $p,
p'\in\G^{r+d}$.
Then $\req_p^{r+d}[i] = \req_{p'}^{r+d}[i]$ holds for all~$i$ in the
range $d< i\le t$.
\end{lemma}
\begin{proof}
We prove the claim by induction on $d$.
The base case is $d=0$, in which round $r+d=r$ is a clean round, and
all non-faulty processes receive the same set of messages.
Thus, by \lineref{3}, we have that
$\req_p^r[i]=\req_{p'}^r[i]$ for all $i$ in the range $d=0 < i \leq t$.
Let $0<d\le t$, and assume inductively that the claim holds for
$d-1$. The inductive assumption guarantees that when the $\req_q$
arrays are sent in round~$r+d$ they agree for all $i$ satisfying
$d-1<i\le t$. In particular, $\max_q \{\req_q[i-1]\}$ is the same for
all $i>d$.
Since $\req_p[i]$ is set to $\max_q \{\req_q[i-1]\}$ on \lineref{3},
it follows that  $\req_p^{r+d}[i] = \req_{p'}^{r+d}[i]$ holds for all
$d<i\le t$, as claimed.
\qed\end{proof}

\hide{\ezra{see if this lemma is actually used anywhere}
\begin{lemma}
$1 \leq \horz_p^k \leq t+1$ holds for all $p\in\p$ and $k\ge 1$.
\end{lemma}
\begin{proof}
The claim is immediate from \lineref{7}, given that the sets $\fail$
and $\failp$ cannot contain more than~$t$ processes.
\qed\end{proof}
}

The purpose of \lineref{7} is to perform our first consistency check,
comparing the reported $\fail_q$ values (from the previous round) to
failures directly observed by~$p$ in the current round (stored in $\fail_p$).
We now show that this can matter only at times $k\le 1$.
At all times~$k\ge 2$, \lineref{7} can be viewed as
having the simpler form of setting the horizon to $t+1-|\failp_p|$.

\begin{lemma}\label{lemma:line7notneeded}
$\horz_p^k = t+1-|\failp_p^k|$ holds after \lineref{7} is executed,
for all times $k\ge 2$ and $p\in\G^k$.
\end{lemma}
\begin{proof}
  If $k\geq 2$ then $k-1 \ge 1$, and so the values of
  $\failp$ received by $p$ at time $k$ contain
  only processes that were indeed faulty by the end of round $k-1$. Since
  failure patterns are monotone, none of these processes sends~$p$
a message in round~$k$. Hence, by \lineref{5} we obtain that
 $\failp_p^k \subseteq \fail_p^k$.
\qed\end{proof}

We denote the first clean round in an execution of \fireAlg by
$\cleanr$. By definition, $\cleanr \geq 1$. We can show:

\begin{lemma}\label{lemma:dechorz}
If $k \geq 1$, then
$\horz_{p'}^{k+1} \leq \horz_p^k$ for every $p,p'\in\G^{k+1}$.
Moreover, for  $k \geq \min\{2, \cleanr\}$
$\horz_{p'}^{k+1} \leq \horz_p^k$ for every $p\in\G^k$ and $p'\in\G^{k+1}$.
\end{lemma}
\begin{proof}
We start with the second case of the lemma:
Let $k \geq \min\{2, \cleanr\}$ and let $p\in\G^k,p'\in\G^{k+1}$.
In particular, either $k\ge 2$, or $k=\cleanr=1$. We consider each of
these cases separately.
Assume  that  $k \geq 2$, and let~$q\in\p$ be a process that updates
$\fail_q$ at time $k-1$. According to
  \lineref{5}, $\fail_q$ contains processes that $q$ does not receive
  messages from during round $k-1$. All of these processes do in fact
  fail no later than round $k-1$.
  Thus, the set $\failp$ computed by process
  $p$ at time $k$ contains only faulty processes. The set $\fail_q$ at
  time $k$ contains all processes of $\fail_q^{k-1}$. Thus, the set
  $\failp_{p'}$ at time $k+1$ contains all processes from
  ${\failp}_p^k$. Hence, ${\failp}_p^k \subseteq
  {\failp}_{p'}^{k+1}$. Therefore, by \lemmaref{lemma:line7notneeded},
  following \lineref{7} by~$p'$ at time $k+1$ we have that
  $\horz_{p'}^{k+1} \leq \horz_p^k$.

  Now consider the case $k=\cleanr=1$. Thus, $p$ and~$p'$ receive the
  same set of   messages during round 1,  and compute $\fail$ and
  $\failp$ in the same manner. Thus,
  $\horz_p^1=\horz_{p'}^1$. Moreover, by \lineref{4} we have that
  $\failp_{p'}^2\supseteq \fail_{p'}^1$. It follows that
  % $|\failp_{p'}^2| \geq    |\fail_p^1|$,
%   and so
  $\min\{|\fail_p^1|, |\failp_p^1|\} =\min\{|\fail_{p'}^1|, |\failp_{p'}^1|\} \leq  |\failp_{p'}^2|$.
By \lemmaref{lemma:line7notneeded} $\horz_{p'}^2 := t+1-|\failp_{p'}^2|$,
hence $\horz_{p'}^2 \leq t+1- \min\{|\fail_p^1|, |\failp_p^1|\} = \horz_p^1$.
 That is, we obtain that $\horz_{p'}^{k+1} \leq \horz_p^k$.

To finish the proof, we are left to handle the case when $k = 1$ and $p,p'\in\G^{2}$.
Since $p \in \G^2$ by time $2$ we have that $p'$ received $p$'s round $2$ messages.
Implying that $\fail_p^1 \subseteq \failp_{p'}^2$. Moreover, due to the monotonicity
of crashes, also $\fail_p^1 \subseteq \fail_{p'}^2$. Therefore,
$\min\{|\failp_{p'}^2|,|\fail_{p'}^2|\} \geq |\fail_p^1| \geq \min\{|\failp_{p}^1|,|\fail_{p}^1|\}$. Hence, by \lineref{7}
$\horz_{p'}^2 \leq \horz_p^1$.
\qed\end{proof}

Denote by $\minH(\F, k)$ the lowest value of $\horz_p^k$, \ie
$\minH(\F, k) = \min_p \{\horz_p^k\}$. When $\F$
is clear from the context, we write $\minH(k)$. Notice that $\minH$ is
the equivalent
of $\rh$ with respect to \fireAlg (recall that~$\rh$ is computed
according to $\ccfs$).

\begin{lemma}\label{lemma:line9doesnot}
Let $k \geq \min\{2,\cleanr\}+1$ and let $p\in\G^k$.
Then, for all  $0   \leq i < t$,   \lineref{9} does not change the
value of $\view_p^k[i]$.
\end{lemma}
\begin{proof}
  Since $k \geq \min\{2,\cleanr\}+1$ we have that $k-1 \geq \min\{2,\cleanr\} \geq 1$.
  At time $k-1$, for every process $q$ and every $0 \leq i \leq t$ it
  holds that $\view_q^{k-1}[i] \geq \horz_q^{k-1}-i$, due to \lineref{9}.
  At time $k$ all processes update $\view$
  according to \lineref{6}, thus setting every entry $i$  (for $i\neq
  t$) to be $\geq \minH(k-1)-i$. By \lemmaref{lemma:dechorz} (recall
  that $k-1 \geq \min\{2,\cleanr\}$) it holds
  that $\max_q\{\horz_q^k\} \leq \minH(k-1)$. Since for every $i \neq t$, $\view_p[i] \geq \minH(k-1)-i$
  it also holds that $\view_p[i] \geq \horz_p-i$. Hence, $\max\{\view_p[i], \horz_p-i\}=\view_p[i]$.
  Thus, for all entries that are not $t$, \lineref{9}
  does not change $\view$.
\qed\end{proof}

\begin{observation}\label{obs:xk}
  For every $k \geq 1$, it holds that $\x(k-1) \leq |\fail_p^{\ \!\!k}| \leq
  \x(k)$. In a similar manner, $\x(k-1) \leq |{\failp}_p^{k+1}| \leq
  \x(k)$.
\end{observation}

\begin{lemma}\label{lemma:firetogether}
  Let $k \geq   \cleanr$ and let $p,p'\in\G^k$.
If $\view_p^k[0]=\view_{p'}^k[0]$ then $p$ and
  $p'$ have the same external output at time~$k$
(\ie they either both fire or they both do not fire at time~$k$).
\end{lemma}
\begin{proof}
  Consider the value of $\view_p^k[0]$. Let $k' \leq k$ be the maximal
  time at
  which $\view_p^k[k-k']$ was updated due to \lineref{8}. Notice that
  $\view_p^{k'}[k-k']=1$, and
  by the update in \lineref{6} it holds that $\view_p^k[0] \geq
  k-k'+1$. Moreover, $\horz_p^{k'}=k-k'+1$, \ie
  $t+1-|{\failp}_p^{k'}|=k-k'+1$.

  Between time $k'-1$ and time $k$ there are $k-k'+1$ rounds. From the above discussion, at time $k'-1$ there were at least $t+k'-k$ failed processes. Thus, between time $k'-1$ and time $k$ there was some clean round. Denote this clean round by $r$.

  By \lemmaref{lemma:oldreq}, for every $i, k-r < i \leq t$, it holds
  that $\req_p^{k}[i] = \req_{p'}^k[i]$. Since
  $\view_p^k[0] \geq k-k'+1 \geq k-r+1$, we have that for every $i,
  \view_p^k[0] \leq i \leq t$ it holds that $\req_p^{k}[i] =
  \req_{p'}^k[i]$. Thus, $p$ and $p'$ either both pass the condition
  of \lineref{10} or they both do not pass. Leading to the fact that
  either $p,p'$ both fire, or they both do not fire.
\qed\end{proof}

\begin{lemma}\label{lemma:zkeq1}
  For every $k \geq \min\{2,\cleanr\}$ and $p\in\G^k$, if
  $\minH(k)=1$ then $\view_p^k[0] = 1$.
\end{lemma}
\begin{proof}
  If $k=\cleanr$ then by \lemmaref{lem:allsame} every process $p$ has
  $\horz_p^k=\minH(k)$. Therefore, if $\minH(k)=1$ then by \lineref{8}, $p$ sets $\view_p^k[0]=1$.

  Continue with the case that $k \neq \cleanr$, \ie $k \geq 2$.
  If $\minH(k)=1$, then some process $q$ has $\horz_q^k=1$. Thus $q$ has
  $|{\failp}_q^k|=t$. Notice that ${\failp}_q^k$ contains processes
  that were faulty during round $k-1$. Therefore, $\fail_q^k =
  {\failp}_q^k$, which leads to the conclusion that all
  $\fail_{q'}^{k-1}$ sets received by $q$ and used in the construction
  of ${\failp}_q^k$ were received from non-faulty processes. Thus, all
  processes receive these sets, and process $p$ also has
  $|{\failp}_{p}^k|=t$ leading to $\horz_{p}^k = 1$. Thus, by
  \lineref{8}, $p$ has $\view_p^k[0]=1$.
\qed\end{proof}

\begin{lemma}\label{lemma:foralli}
  Let $r \geq \cleanr$ and  $p,p'\in\G^r$.
  If $\minH(r) > 1$   then $\view^r_p[i] =  \view^r_{p'}[i]$ holds
  for all $0 \leq i < \minH(r)-1$.
\end{lemma}
\begin{proof}
  The proof is by induction on~$r\ge \cleanr$.
  For $r=\cleanr$,   we have by \lemmaref{lem:allsame} that
$\view_p =  \view_{p'}$, and the claim immediately follows.
  For the inductive step, assume that $r>\cleanr$ and that the claim
  holds for $r-1$.
  We consider two cases.   First assume that $\minH(r)=\minH(r-1)$.
  In this case, no process failure is discovered in round~$r$. Thus,
  round~$r$ is clean, and the claim follows by \lemmaref{lem:allsame}
  as in the base case.

Next, assume that  $\minH(r) < \minH(r-1)$.
The $\view_p[i]$ values can change only on \lineref{6}, \lineref{8},
and \lineref{9}. First consider the change by \lineref{6}.
In this case, \view$[i-1]$ is set to $\min_q\{\view_q[i]\}+1$ for
$1 \leq i \leq t$. By the inductive assumption we have that
$\view_p[i] =  \view_{p'}[i]$ holds for all $0 \leq i < \minH(r)-1$
before \lineref{6} is applied. Since the values of $\view[j]$ before
\lineref{6} are shifted down by one, and become the values of
$\view[j-1]$ after it is applied, we obtain that $\view_p[i] = \view_{p'}[i]$
for all $0 \leq i < \minH(r-1)-2$ once \lineref{6} has completed.
Since $\minH(r)<\minH(r-1)$, we have that
$\minH(r)-1\le\minH(r-1)-2$. Consequently,  $\view_p[i] = \view_{p'}[i]$
for all $0 \leq i < \minH(r)-1$ when \lineref{7} is reached.

On \lineref{8}, $\view_p[\horz_p^r-1]$ is set to~1.
By definition, $\minH(r)\le \horz_p^r$, so the update does not affect
values \view$[i]$ for $i<\minH(r)-1$. Hence, the fact that
$\view_p[i] = \view_{p'}[i]$
for all $0 \leq i < \minH(r)-1$, which was shown above to hold
when \lineref{7} is reached also holds when \lineref{9} is reached.

By \lemmaref{lemma:line9doesnot}, since $r-1 \geq \cleanr$ \lineref{9}
does not change the value of $\view_p^r[i]$, for all $0 \leq i < t$. Since $\minH(r)-1 \leq t$
we have that after \lineref{9} $\view_p[i] = \view_{p'}[i]$ for all $0 \leq i < \minH(r)-1$.
\qed\end{proof}

\begin{lemma}\label{lemma:simul}
\simul holds for all times $k \geq \cleanr$.
\end{lemma}
\begin{proof}
  By \lemmaref{lemma:zkeq1} and \lemmaref{lemma:foralli}, for two processes $p,p'$ it holds that $\view_p[0] = \view_{p'}[0]$. Together with \lemmaref{lemma:firetogether} we have that $p,p'$ fire together or do not fire together, for every $r \geq \cleanr$.
\qed\end{proof}

\begin{lemma}\label{lemma:liveness}
\liveness holds for all times~$k \geq 0$.
\end{lemma}
\begin{proof}
  If some non-faulty process $p$ received a request to fire at time $k$, then it sets $\req_p^k[0]=1$.
  Since $\view_p^k[0] \geq \horz_p^k \geq 1$, $p$ will not update $\req_p^k[0]=0$ due to \lineref{11}.
  Thus, at time $k+1$ it holds that $\req_p^{k+1}[1]=1$; and in general, if by time $k+i$ $p$ does not set $\req_p^{k+i}[i]=0$ then it holds that $\req_p^{k+i+1}[i+1]=1$.

  Notice that if $p$ sets $\req_p^{k+i}[i]=0$ (for $i\geq 1$), then
  $p$ executes \lineref{11}, indicating that $p$ fires.
  Notice that $\view_p^{k+t+1}[0] \leq \horz_p^{k+t+1} \leq t+1$. Thus, if by time $k+t+1$ $p$ has not set $\req_p^{k+t+1}[t+1]=0$, then at time $k+t+1$ $p$ will fire.

  And we conclude that within $t+1$ rounds $p$ will fire, and \liveness holds.
\qed\end{proof}

\begin{lemma}\label{lemma:viewhorz}
  Let $k \geq 1$, and let $p\in\p$.
  If $k'$ is such that $p\in\G^{k'}$ and $k'=k+\horz_p^k-1$,
  then $\view_p^{k'}[0] \leq \horz_p^k$.
\end{lemma}
\begin{proof}
  Let $p$ be any process and consider time $k$: by \lineref{8} process $p$ sets $\view_p^{k}[\horz_p^k-1]=1$.
  For $\horz_p^k=1$, it holds that $\view_p^{k}[\horz_p^k-1]=\view_p^{k}[0]=1 \leq \horz_p^k$.

  The rest of the proof concentrates on the case that $\horz_p^k > 1$. At time $k'=k+1$ if $p$ does not update $\view_p^{k+1}[\horz_p^k-2]$ due to \lineref{8}, it holds that $\view_p^{k+1}[\horz_p^k-2] \leq 2$; and in general, if at time $k'=k+j$ $p$ does not update $\view_p^{k+j}[\horz_p^k-1-j]$ then $\view_p^{k+j}[\horz_p^k-1-j] \leq 1 + j$. Notice that if $p$ does update $\view_p[\horz_p^k-1-j]$ due to \lineref{8} then $p$ has $\view_p[\horz_p^k-1-j] =1 \leq 1 + j$.

  Thus, at time $k'=k+\horz_p^k-1$ it holds that $\view_p^{k'}[0] \leq \horz_p^k$.
\qed\end{proof}

Define
$\minHG(\F,k)=\min_{p\in\G}\horz_p^{k}$ and use it to define
$\ZH(\F, k) = \min_{k' \geq k} \{k'+ \minHG(\F,k'+1) \}$. If $\F$ is
clear from the context, we use $\ZH(k)$.

Notice that $\minHG$ is similar to $\minH$
except that $\minHG$ considers only $\horz$ values of processes that never
crash, while
$\minH$ considers processes that haven't crashed yet. Also, notice that $\ZH$ is
the equivalent of $\bb$ with respect to $\fireAlg$ (recall that $\bb$
is computed according to $\ccfs$).

\begin{lemma}\label{lemma:zklebk}
  $\ZH(k) \leq \bb(k)$, for every $k \geq 0$.
\end{lemma}
\begin{proof}
  Consider the value of $\bb(k) = \min_{k' \geq k}\{\ah(k')\}$, and denote by $k''$ the
  latest time for which the minimum is reached. \Ie $\bb(k) = \ah(k'') = k''+t+1-x(k'')$, and for all $k' > k''$ it holds that $\ah(k') > \bb(k)$. Thus, $\x(k''+1) = \x(k'')$ (otherwise, $\ah(k''+1) \leq \ah(k'')$,
  contradicting the choice of $k''$).

  Since $\x(k''+1) = \x(k'')$ it holds that no new failed processes are discovered at round $k''+1$.
  Consider two options, $k'' \geq 1$ and $k''=0$. When $k'' \geq 1$ it follows that $k''+1 \geq 2$ and therefore
  every non-faulty process $p$ at time $k''+1$ has $\horz_p^{k''+1}=t+1-\x(k'')$. Thus, $\minHG(k''+1)=t+1-\x(k'')$ leading to $k''+\minHG(k''+1)=k''+\rh(k'') = \ah(k'')$.

  Consider the case that $k''=0$. By \definitionref{def:x}, $\x(k'')=\x(0)=0$ leading to $\rh(k'')=t+1$. Since $\horz_p^1 \leq t+1$ it
  follows that $k'' + \minHG(k''+1) \leq k'' + \rh(k'')=\ah(ak'')$.

  For both $k'' \geq 1$ and $k''=0$ we conclude that $k'' + \minHG(k''+1) \leq \ah(ak'')$.
  Since $\bb(k)=\ah(k'')$ we conclude that $\ZH(k) \leq \bb(k)$.
\qed\end{proof}

\begin{lemma}\label{lemma:zkleqk}
  \safety holds at all times $k \geq \bb(0)$.
\end{lemma}
\begin{proof}
  Let $p \in \G$ be a
  process such that $\horz_p^{k'+1}=\minHG(k'+1)$.
  By \lemmaref{lemma:dechorz}, for all $k'' \geq k'+1$ it holds that
  $\horz_p^{k''} \leq \horz_p^{k'+1}$ (notice that $k'+1 \geq 1$, and $p \in \G^{k''}$).

  Consider time $k'+i$ (for $i \geq 1$), by \lemmaref{lemma:viewhorz} for every time $k'' = k'+i+\horz_p^{k'+i}-1$ it holds
  that $\view_p^{k''}[0] \leq \horz_p^{k'+i} \leq \horz_p^{k'+1}$. Thus,
  for every time $k'' \geq k'+\horz_p^{k'+1}=\ZH(0)$ it holds that $\view_p^{k''}[0] \leq \horz_p^{k'+1} \leq \ZH(0)$.

  By \lemmaref{lemma:zklebk} we have that $\ZH(0) \leq \bb(0)$. Hence,
  For every time $k'' \geq \bb(0)$ it holds that $\view_p^{k''}[0] \leq \bb(0)$.
  Consider time $\bb(0)$. Since $\cleanr \leq \bb(0)$, by \lemmaref{lemma:simul}, \simul holds.
  Therefore, if some process fires then all processes in $\G^{\bb(0)}$ fire. For
  any process $q \in \G^{\bb(0)}$. If $q$ fires, then it sets all $\req_q^{\bb(0)}[i]=0$
  for all $i \geq \view_q^{\bb(0)}[0]$. If $q$ does not fire, then it is because
  $\req_q^{\bb(0)}[i]=0$ for all $i \geq \view_q^{\bb(0)}[0]$. Moreover,
  since $\view_q^{\bb(0)}[0] \leq \bb(0)$, it holds that $\req_q^{\bb(0)}[i]=0$ for all $i \geq \bb(0)$.

  Since for every $k'' \geq {\bb(0)}$ it holds that $\view_p^{k''}[0] \leq
  \bb(0)$, we have that if process $p$ has $\req_p^k[i]=1$, it
  must have been set at some time $\geq 0$. In other words, if a fire
  action occurs then there was a previous $\go$ input received; and
  because $\req_p^k[i]$ is zeroed once a fire action occurs, each
  $\go$ can induce at most a single fire action. Thus, the number of
  times $\ZH(0) \leq k' \leq k$ for which a fire action occurs is not
  larger than
  the number of times $0 \leq k' < k$ during which a $\go$ input is received.
\qed\end{proof}

\begin{lemma}\label{lemma:firespeed}
  Let input $\I$ be sequential with respect to $(\fireAlg, \s, \F)$.
  If $\I_p^k=1$ for process $p$ at time $k$ then $\OO_{p}^{k'}=1$ for $k<k' \leq \ZH(k)$.
\end{lemma}
\begin{proof}
  Since $\I$ is sequential and $\I_p^k=1$ it holds that $p \in \G$.
  Consider $\ZH(0) =  \min_{i} \{i + \minHG(i+1)\}$, and denote by $k'$ a time
  that satisfies $k'+\minHG(k'+1) = \ZH(0)$. Let $q \in \G$ be some process
  such that $\horz_q^{k'+1}=\minHG(k'+1)$.
  Since $k'+1 \geq 1$ and $q \in \G$,  by \lemmaref{lemma:viewhorz}, at
  time $k''=k'+\horz_q^{k'+1}=\ZH(0)$ it holds
  that $\view_q^{k''}[0] \leq \horz_q^{k'+1}$.

  If $p$ fires at some time $k < k'' < \ZH(k)$ then the claim is proved.
  Otherwise, at time $k''=\ZH(k)$ it holds that
  $\view_q^{k''}[0] \leq \horz_q^{k'+1}$. Since $\I_k^p=1$ and $p\in\G$, by time $k+1$ we have
  that $\req_q^{k+1}[1]=1$. Since $p$ does not fire before time $\ZH(k)$ and since
  \simul holds, we have that by time $k''=\ZH(k)$ it holds that $\req_q^{k''}[\horz_q^{k'+1}]=1$.
  Therefore, at time $k''=\ZH(k)$
  $q$ will fire and due to \simul $p$ will fire as well.
  And we conclude that for some time $k''$, satisfying $k < k'' \leq \ZH(k)$,
  we have that $\OO_p^{k''}=1$.
\qed\end{proof}

\begin{theorem}\label{thm:ssfs}
  \fireAlg solves the SSFS problem, it optimally stabilizes and is optimally swift.
\end{theorem}
\begin{proof}
  Consider any initial state $\s$, any input path $\I$ and any failure
  pattern $\F$.
  By definition, $\bb(\F, 0) \leq t+1$. Thus, by
  \lemmaref{lemma:zkleqk}, \safety holds starting from time
  $t+1$. Since by time $t+1$ there is
  a clean round, by \lemmaref{lemma:simul}, \simul holds starting from
  time $t+1$. \lemmaref{lemma:liveness}
  finishes the proof, and we have that for time $k=t+1$ it holds that
  $\stab(\fireAlg, \s, \I, \F) \leq k$.
\qed\end{proof}

\begin{theorem}\label{thm:stabotimal}
\fireAlg optimally stabilizes.
\end{theorem}
\begin{proof}
  By \lemmaref{lemma:zkleqk}, the \safety property of \fireAlg holds from
  time $\bb(\F,0)$. Moreover, by \lemmaref{lemma:simul} together with
  the fact that by time $\bb(\F,0)$ there is a clean round, the
  \simul property of \fireAlg holds from time $\bb(\F,0)$. Combined
  with \lemmaref{lemma:liveness} we have
  that $\stab(\fireAlg, \s, \I, \F) \leq \bb(\F,0)$; for any state
  $\s$, input path $\I$ and failure pattern $\F$.
  \Ie  $\max_{\s, \I}\{\stab(\fireAlg,\s,\I,\F)\} \leq \bb(\F, 0)$.

  Let $\A$ be any SSFS algorithm. By \theoremref{theorem:lowerbound1} for
  every failure pattern $\F$ we have that $\max_{\s,
    \I}\{\stab(\A,\s,\I,\F)\} \geq \bb(\F, 0)$. Thus, for every $\F$:
  $\max_{\s, \I}\{\stab(\fireAlg,\s,\I,\F)\} \leq \max_{\s,
    \I}\{\stab(\A,\s,\I,\F)\}$.
\qed\end{proof}

\begin{theorem}\label{thm:quickoptimal}
  \fireAlg is optimally swift.
\end{theorem}
\begin{proof}
  Let input $\I$ be sequential with respect to $(\fireAlg, \s_\fireAlg, \F)$.
  By \lemmaref{lemma:firespeed}, if $\I_p^k=1$ for some process $p$ at
  time $k$ then for some $k'$ satisfying $k < k'' \leq \ZH(k)$ it holds
  that $\OO_{p}^{k'}=1$. Therefore, by time $\ZH(k)$ we
  have that $\#[(\fireAlg,\s_\fireAlg,\I,\F),\ZH(k)]$
  is no smaller than the number of $\go$ inputs received by time $k$.

  Let $\A$ be any SSFS algorithm and $\I$ sequential with respect to
  $(\A, \s_\A, \F)$. By \theoremref{theorem:lowerbound2}, for every $k
  \geq 0$ for which a $\go$ input is received in $\I^k$ there is no
  fire action in $\OO\!=\!\A(\s_\A, \I, \F)$ during
  times $k'$ satisfying $k < k' < \bb(\F, k)$. Since $\ZH(k) \leq
  \bb(k)$ (\lemmaref{lemma:zklebk}), it
  holds that by time $\ZH(k)$, the value of $\#[(\A,\s_\A,\I,\F),\ZH(k)]$
  is at most equal to the number of $\go$ inputs received by time $k$.

  Thus, for every $\s_\A, \s_\fireAlg,\F$ and sequential $\I$ it holds that\\
  $\#[(\fireAlg,\s_\fireAlg,\I,\F),k] \geq \#[(\A,\s_\A,\I,\F),k]$, for all $k$.
%  That is, $\fireAlg \leq_\quick \A$.
\qed\end{proof}

\section{Conclusions and Open Problems}\label{sec:discussion}
This paper presents \fireAlg, the first self-stabilizing
firing squad algorithm.
\fireAlg is optimal in two important respects:
It optimally stabilizes, and is optimally swift.
There are many directions in which this work can be extended.
These include:
\begin{itemize}
  \item \fireAlg assumes the crash fault model. What can be said about
    the omission fault model? And what about the \byzantine fault
    model? Each such extension seems to be a nontrivial step.
\item \fireAlg works when we assume that failures are permanent. Being
  an ongoing and everlasting service, firing squad is expected to
  operate for long periods, in which processes may recover.
A more reasonable assumption in this case is that there is a bound
(of~$t$) on
the number of failures over every interval of~$m$ rounds, for
some~$m$. (Non-stabilizing) Continuous consensus has recently been
studied in this model \cite{MM2}, and it would be interesting to see
if the same can  be done for self-stabilizing firing squad.
\end{itemize}

\bibliographystyle{plain}
\bibliography{bibliography}
\end{document}